\documentclass[12pt]{article}
\usepackage{amsmath,amssymb,amsthm,amsxtra,bbm,overpic,bm,epsfig,subfigure}
\usepackage{hyperref}
\usepackage{mathrsfs}
\usepackage{enumitem}
\usepackage{graphicx}
\usepackage{color}
\usepackage{comment}
\usepackage{epstopdf}
\usepackage{float}
\usepackage{cite}
\allowdisplaybreaks[2]

\usepackage{slashed,stmaryrd,multirow}

\def\thefootnote{\fnsymbol{footnote}}

\usepackage{hyperref}

\usepackage{lscape}%
\usepackage{array}
\usepackage{booktabs}%

\begin{document}

\vspace{0.2cm}

\begin{center}
{\large\bf Towards a detection of reactor $\overline{\nu}^{}_e \to \overline{\nu}^{}_\mu$ 
and $\overline{\nu}^{}_e \to \overline{\nu}^{}_\tau$ oscillations \\
with possible CP violation}
\end{center}

\vspace{0.2cm}

\begin{center}
{\bf Yifang Wang}~\footnote{E-mail: yfwang@ihep.ac.cn} , \;
{\bf Zhi-zhong Xing}~\footnote{E-mail: xingzz@ihep.ac.cn} , \;
{\bf Shun Zhou}~\footnote{E-mail: zhoush@ihep.ac.cn}
\\
\vspace{0.2cm}
{\small Institute of High Energy Physics, Chinese Academy of Sciences, Beijing 100049, China\\
University of Chinese Academy of Sciences, Beijing 100049, China}
\end{center}

\vspace{1.5cm}

\begin{abstract}
We propose an unprecedented detection of reactor $\overline{\nu}^{}_e \to
\overline{\nu}^{}_\mu$ and $\overline{\nu}^{}_e \to \overline{\nu}^{}_\tau$
oscillations by using elastic antineutrino-electron scattering processes
$\overline{\nu}^{}_\alpha + e^- \to \overline{\nu}^{}_\alpha + e^-$ (for
$\alpha = e, \mu, \tau$), among which the $\overline{\nu}^{}_e$ events can be
singled out by accurately measuring the $\overline{\nu}^{}_e$ flux via the
inverse beta decay $\overline{\nu}^{}_e + p \to e^+ + n$. A proof-of-concept
study shows that such measurements will not only be able to test the
conservation of probability for reactor antineutrino oscillations, but also
offer a new possibility to probe leptonic CP violation at the one-loop level.
\end{abstract}

\newpage

\def\thefootnote{\arabic{footnote}}
\setcounter{footnote}{0}

\section{Introduction}

The reactor-based neutrino experiments have played a crucial role in the
developments of nuclear and particle physics~\cite{Wang:2014tpa, Wen:2017hbg}, 
such as the discovery of the electron antineutrino 
$\overline{\nu}^{}_e$~\cite{Cowan:1956rrn}, the discoveries of long- and 
short-baseline $\overline{\nu}^{}_e \to \overline{\nu}^{}_e$
oscillations~\cite{KamLAND:2002uet,DayaBay:2012fng}, and the first
measurement of the smallest lepton flavor mixing
angles $\theta^{}_{13}$~\cite{DayaBay:2012fng,RENO:2012mkc}.
The new flagship reactor neutrino oscillation experiment
JUNO~\cite{JUNO:2015zny,JUNO:2021vlw}, a medium-baseline facility which has just
started data taking, aims to resolve another fundamental issue in particle
physics and cosmology --- the mass ordering of three active neutrinos.

It is well known that the flavor oscillations of reactor antineutrinos
belong to the {\it disappearance} category, in the sense that the
$\overline{\nu}^{}_e$ events observed at the far detector are somewhat
fewer than those recorded at the near detector. Although
$\overline{\nu}^{}_e \to \overline{\nu}^{}_\mu$ and
$\overline{\nu}^{}_e \to \overline{\nu}^{}_\tau$ oscillations {\it do} take place
in a reactor experiment, they cannot be directly detected via the corresponding weak
{\it charged-current} interactions associated with $\overline{\nu}^{}_\mu$ and
$\overline{\nu}^{}_\tau$~\cite{Xing:2011zza}. The reason is simply that the reactor
antineutrino beam energy is too low to produce the $\mu^+$ or $\tau^+$ events
via the $\overline{\nu}^{}_\mu + p \to \mu^+ + n$ or
$\overline{\nu}^{}_\tau + p \to \tau^+ + n$ processes in the detector. A
burning question is therefore whether there exists a way out of this impasse
in the coming precision measurement era.

The answer to this important question will be affirmative,
if a dedicated measurement of the appearance
of $\overline{\nu}^{}_\mu$ and $\overline{\nu}^{}_\tau$ events can be done
by means of their weak {\it neutral-current} interactions with the target
material. In this note we are going to propose an unprecedented detection of
the reactor $\overline{\nu}^{}_e \to \overline{\nu}^{}_\mu$ and
$\overline{\nu}^{}_e \to \overline{\nu}^{}_\tau$ oscillations with the help
of elastic antineutrino-electron scattering processes
$\overline{\nu}^{}_\alpha + e^- \to \overline{\nu}^{}_\alpha + e^-$ (for
$\alpha = e, \mu, \tau$), among which the $\overline{\nu}^{}_e$ events can be
singled out by detecting the $\overline{\nu}^{}_e$ flux via the inverse beta
decay $\overline{\nu}^{}_e + p \to e^+ + n$ to a sufficiently high degree
of accuracy. We find that such precision measurements will help
open a new window for experimental neutrino physics at least in the following
three aspects:
\begin{enumerate}
\item     to directly confirm the appearance of $\overline{\nu}^{}_\mu$ and
$\overline{\nu}^{}_\tau$ events originating from the initial
$\overline{\nu}^{}_e$ events of a nuclear reactor via flavor oscillations,
and thus to test the conservation of probability for reactor antineutrino
oscillations by combining it with the
$\overline{\nu}^{}_e \to \overline{\nu}^{}_e$ disappearance.

\item     to probe a fine difference between
$\overline{\nu}^{}_e \to \overline{\nu}^{}_\mu$ and
$\overline{\nu}^{}_e \to \overline{\nu}^{}_\tau$ oscillations, which is sensitive
to both the $\mu$-$\tau$ interchange symmetry and the leptonic CP violation, by
precisely measuring the cross sections of elastic
$\overline{\nu}^{}_\mu$-$e^-$ and $\overline{\nu}^{}_\tau$-$e^-$ scattering
reactions at the {\it one-loop} level.

\item     to search for possible new physics either beyond the standard weak
interactions or beyond the standard three-flavor oscillation scheme, or both
of them.
\end{enumerate}
We expect that the experimental and theoretical studies of this kind will
find more applications at the low-energy luminosity or intensity frontiers
of particle physics.

The present work is intended to provide a {\it proof-of-concept} investigation
of points 1 and 2 listed above, with a very preliminary numerical illustration
by taking the JUNO experiment for example. In particular, we highlight the novel
possibility of probing or constraining the leptonic CP-violating phase from a
precision measurement of the reactor $\overline{\nu}^{}_e \to \overline{\nu}^{}_\mu$
and $\overline{\nu}^{}_e \to \overline{\nu}^{}_\tau$ oscillations.

\section{Flavor oscillations}

Let us focus on the standard three-flavor oscillation scheme, in which the
$3\times 3$ unitary Pontecorvo-Maki-Nakagawa-Sakata (PMNS) neutrino mixing
matrix $U$~\cite{Pontecorvo:1957cp,Maki:1962mu} can be parameterized as
\footnote{Possible non-unitarity of $U$ has been constrained to be below
${\cal O}\left(10^{-3}\right)$ in the canonical seesaw
mechanism~\cite{Blennow:2023mqx}, and thus can be neglected in this work.
Here the Majorana phases of $U$ are not taken into account either, as they
are insensitive to the reactor antineutrino oscillations under discussion.}
\begin{equation}
U = \left( \begin{matrix} U^{}_{e 1} & U^{}_{e 2} & U^{}_{e 3} \cr
U^{}_{\mu 1} & U^{}_{\mu 2} & U^{}_{\mu 3} \cr
U^{}_{\tau 1} & U^{}_{\tau 2} & U^{}_{\tau 3} \end{matrix} \right)
= \left( \begin{matrix}
c^{}_{12} c^{}_{13} & s^{}_{12} c^{}_{13} & \hat{s}^{*}_{13} \\
-s^{}_{12} c^{}_{23} - c^{}_{12} \hat{s}^{}_{13} s^{}_{23}
& c^{}_{12} c^{}_{23} - s^{}_{12} \hat{s}^{}_{13} s^{}_{23}
& c^{}_{13} s^{}_{23} \\
s^{}_{12} s^{}_{23} - c^{}_{12} \hat{s}^{}_{13} c^{}_{23}
& ~ -c^{}_{12} s^{}_{23} - s^{}_{12} \hat{s}^{}_{13} c^{}_{23} ~
& c^{}_{13} c^{}_{23}
\end{matrix}\right) \; ,
\label{1}
\end{equation}
where $c^{}_{ij} \equiv \cos \theta^{}_{ij}$, $s^{}_{ij} \equiv \sin \theta^{}_{ij}$
and $\hat{s}^{}_{13} \equiv s^{}_{13} e^{{\rm i}\delta}$
with $\theta^{}_{ij}$ (for $ij = 12, 13, 23$) being the flavor mixing
angles and $\delta$ being the nontrivial phase responsible for leptonic CP violation
in neutrino oscillations. The probabilities of reactor
$\overline{\nu}^{}_e \to \overline{\nu}^{}_\alpha$ oscillations in vacuum are given by
\begin{eqnarray}
P(\overline{\nu}^{}_e \to \overline{\nu}^{}_\alpha) &=& \delta^{}_{e\alpha}
- 4\sum_{i<j} {\rm Re}\left( U^{}_{ei} U^*_{ej} U^*_{\alpha i} U^{}_{\alpha j}\right)
\sin^2 F^{}_{ji} - 8 {\cal J} \sum_\beta \epsilon^{}_{e\alpha \beta} \prod_{i<j}
\sin F^{}_{ji} \; ,
\label{2}
\end{eqnarray}
where $F^{}_{ji} \equiv \left(m^2_j - m^2_i\right) L/(4E)$ with $m^{}_{i,j}$
being the neutrino masses are defined (for $i, j = 1, 2, 3$),
$E$ represents the average antineutrino beam energy, $L$ denotes the baseline length,
$\epsilon^{}_{e \alpha\beta}$ stands for the three-dimensional Levi-Civita symbol
(for $\alpha, \beta = e, \mu, \tau$), and
\begin{eqnarray}
{\cal J} = \frac{1}{8} \sin 2\theta^{}_{12} \sin 2\theta^{}_{13} \cos\theta^{}_{13}
\sin 2\theta^{}_{23} \sin\delta \;
\label{3}
\end{eqnarray}
is the unqiue Jarlskog invariant of CP violation for the PMNS lepton flavor mixing
matrix~\cite{Jarlskog:1985ht,Wu:1985ea}.

It is obvious that the {\it disappearance} oscillation channel
$\overline{\nu}^{}_e \to \overline{\nu}^{}_e$ is CP-conserving, while the
{\it appearance} oscillation channels $\overline{\nu}^{}_e \to \overline{\nu}^{}_\mu$
and $\overline{\nu}^{}_e \to \overline{\nu}^{}_\tau$ contain the CP-violating terms
of the same magnitude but the opposite signs. That is why we find it useful to redefine
\begin{eqnarray}
P^{}_+ & \equiv & P(\overline{\nu}^{}_e \to \overline{\nu}^{}_\mu) +
P(\overline{\nu}^{}_e \to \overline{\nu}^{}_\tau) = 1 - P(\overline{\nu}^{}_e \to
\overline{\nu}^{}_e) = 4\sum_{i<j} |U^{}_{ei}|^2 |U^{}_{ej}|^2 \sin^2 F^{}_{ji} \; ,
\nonumber \\
P^{}_- & \equiv & P(\overline{\nu}^{}_e \to \overline{\nu}^{}_\mu) -
P(\overline{\nu}^{}_e \to \overline{\nu}^{}_\tau)
 = \sum_{i<j} D^{}_{ij} \sin^2 F^{}_{ji}-16 {\cal J} \prod_{i<j} \sin F^{}_{ji} \; ,
\label{4}
\end{eqnarray}
where $D^{}_{ij} \equiv 4 {\rm Re}\left[ U^{}_{ei} U^*_{ej}
\left( U^*_{\tau i} U^{}_{\tau j} - U^*_{\mu i} U^{}_{\mu j} \right) \right]$
characterize the effects of $\mu$-$\tau$ interchange symmetry breaking for the
PMNS matrix $U$. To be explicit, we obtain
\begin{eqnarray}
D^{}_{12} & = & + \sin^2 2\theta^{}_{12} \cos^2\theta^{}_{13} \cos 2\theta^{}_{23}
\left( 1 + \sin^2\theta^{}_{13} + 2 \cot 2\theta^{}_{12} \sin\theta^{}_{13}
\tan 2\theta^{}_{23} \cos\delta \right) \; ,
\nonumber \\
D^{}_{13} & = & - 2\cos^2\theta^{}_{12} \sin 2\theta^{}_{13} \cos\theta^{}_{13}
\cos 2\theta^{}_{23} \left(\sin\theta^{}_{13} - \tan\theta^{}_{12} \tan 2\theta^{}_{23}
\cos \delta \right) \; ,
\nonumber \\
D^{}_{23} & = & - 2\sin^2\theta^{}_{12} \sin 2\theta^{}_{13} \cos\theta^{}_{13}
\cos 2\theta^{}_{23} \left(\sin\theta^{}_{13} + \cot\theta^{}_{12} \tan 2\theta^{}_{23}
\cos \delta \right) \; ,
\label{5}
\end{eqnarray}
which will vanish in the limit of the $\mu$-$\tau$ permutation symmetry (i.e.,
$\theta^{}_{23} = \pi/4$ and $\theta^{}_{13} = 0$) or the $\mu$-$\tau$ reflection
symmetry (i.e., $\theta^{}_{23} = \pi/4$ and $\delta = \pm \pi/2$)~\cite{Xing:2015fdg}.
Given the facts that the best-fit value of $\theta^{}_{23}$ is really not far away
from $\pi/4$ and the maximum value of $|{\cal J}|$ is about $3\%$ in a global
analysis of current neutrino oscillation data~\cite{Esteban:2024eli,Capozzi:2025wyn},
we conclude that the magnitudes of $D^{}_{ij}$ and ${\cal J}$ are both suppressed,
leading to an unfortunate suppression of $P^{}_-$ as compared with $P^{}_+$.

Note that the reactor $\overline{\nu}^{}_e \to \overline{\nu}^{}_e$ oscillations
have been firmly established by measuring the survival $\overline{\nu}^{}_e$ events
at the far detector with the help of the weak charged-current
$\overline{\nu}^{}_e + p \to e^+ + n$ reactions. The first purpose of this work
is to propose a careful precision measurement of $P^{}_+$ by use of
elastic $\overline{\nu}^{}_\alpha + e^- \to \overline{\nu}^{}_\alpha + e^-$
scattering (for $\alpha = \mu, \tau$) via the weak neutral-current interactions
in a reactor-based experiment like the JUNO, such that one may effectively test the
consistency of the standard three-flavor neutrino oscillation framework by
examining the conservation of probability associated with these pure quantum
processes [i.e., $P^{}_+ + P(\overline{\nu}^{}_e \to \overline{\nu}^{}_e) = 1$
should hold].
\begin{figure}[t!]
\centering
\includegraphics[width=\textwidth]{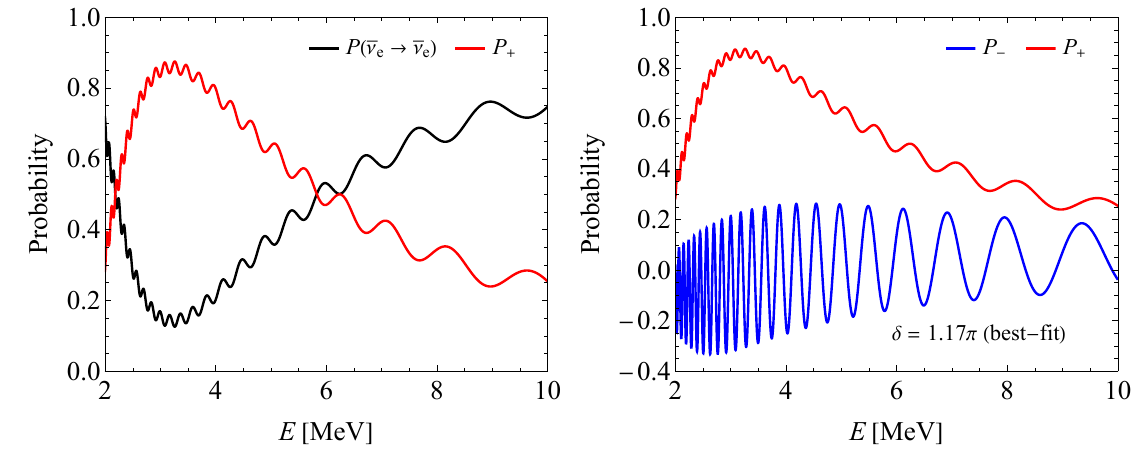}
\vspace{-1cm}
\caption{The sum of the {\it appearance}
oscillation probabilities $P^{}_+ \equiv P(\overline{\nu}^{}_e \to
\overline{\nu}^{}_\mu) + P(\overline{\nu}^{}_e \to \overline{\nu}^{}_\tau)$
versus the survival probability $P(\overline{\nu}^{}_e \to \overline{\nu}^{}_e)$
changing with the beam energy $E$ is shown in the left panel, while $P^{}_+$ 
and $P^{}_- \equiv P(\overline{\nu}^{}_e \to
\overline{\nu}^{}_\mu) - P(\overline{\nu}^{}_e \to \overline{\nu}^{}_\tau)$ 
in the right panel. The best-fit values of the relevant oscillation parameters 
$\sin^2 \theta^{}_{12} = 0.308$, $\sin^2 \theta^{}_{13} = 0.02215$, 
$\sin^2\theta^{}_{23} = 0.470$ and $\delta = 1.17\pi$, together with 
$m^2_2 - m^2_1 = 7.49\times 10^{-5}~{\rm eV}^2$ and 
$m^2_3 - m^2_1 = 2.513\times 10^{-3}~{\rm eV}^2$ (normal
mass ordering) have been input~\cite{Esteban:2024eli,Capozzi:2025wyn}, 
and the baseline length is set to $L = 53~{\rm km}$.}
\label{fig:pplus}
\end{figure}

For the sake of illustration, we show the probabilities $P^{}_+$ and
$P(\overline{\nu}^{}_e \to \overline{\nu}^{}_e)$ changing with the reactor
antineutrino beam energy $E$ in the left panel of Fig.~\ref{fig:pplus},
where $L = 53~{\rm km}$ has been taken as a typical baseline length. 
In addition, the best-fit values of the oscillation parameters 
$\sin^2 \theta^{}_{12} = 0.308$, $\sin^2 \theta^{}_{13} = 0.02215$, 
$\sin^2\theta^{}_{23} = 0.470$ and $\delta = 1.17\pi$, together with 
$m^2_2 - m^2_1 = 7.49\times 10^{-5}~{\rm eV}^2$ and 
$m^2_3 - m^2_1 = 2.513\times 10^{-3}~{\rm eV}^2$ (normal mass ordering) 
are taken from Ref.~\cite{Esteban:2024eli}, where a perfect agreement 
with the results from an independent global-fit analysis of
neutrino oscillation data in Ref.~\cite{Capozzi:2025wyn} can be found. From Fig.~\ref{fig:pplus}, it is clear that a significant fraction of the
$\overline{\nu}^{}_e$ events oscillate into the $\overline{\nu}^{}_\mu$ and
$\overline{\nu}^{}_\tau$ events, especially in the energy range
$E \in [2, 6]~{\rm MeV}$. One may therefore expect that the antineutrinos of all
three flavors are detectable in a liquid-scintillator detector.

In comparison, it is certainly more ambitious and thus more challenging to detect
$P^{}_-$ by means of the same techniques, so as to probe or constrain the highly
nontrivial effects of leptonic CP violation and $\mu$-$\tau$ interchange symmetry
breaking. The right panel of Fig.~\ref{fig:pplus} illustrates how small $P^{}_-$ 
is expected to be
with respect to $P^{}_+$. A delicate measurement of the cross sections of elastic
$\overline{\nu}^{}_\mu$-$e^-$ and $\overline{\nu}^{}_\tau$-$e^-$ scattering at the
one-loop level of accuracy is mandatory, in order to distinguish between the
$\overline{\nu}^{}_\mu$ and $\overline{\nu}^{}_\tau$ events that originate from
the reactor $\overline{\nu}^{}_e \to \overline{\nu}^{}_\mu$
and $\overline{\nu}^{}_e \to \overline{\nu}^{}_\tau$ oscillations.

In view of the fact that the behaviors of reactor antineutrino oscillations in a
medium-baseline experiment are essentially insensitive to terrestrial matter
effects
\footnote{In fact, only $\theta^{}_{12}$ and $F^{}_{21}$ are slightly
contaminated by terrestrial matter effects for a medium-baseline reactor antineutrino
experiment like JUNO~\cite{Li:2016txk,Li:2018jgd,Capozzi:2020cxm}. It is also
known that the combination ``$\sin 2\theta^{}_{23} \sin\delta$" is in particular
insensitive to the matter-induced corrections~\cite{Toshev:1991ku}.},
we shall not take into account such insignificant corrections in the following
proof-of-concept analysis.
\begin{figure}[t!]
\centering
\includegraphics[width=0.55\textwidth]{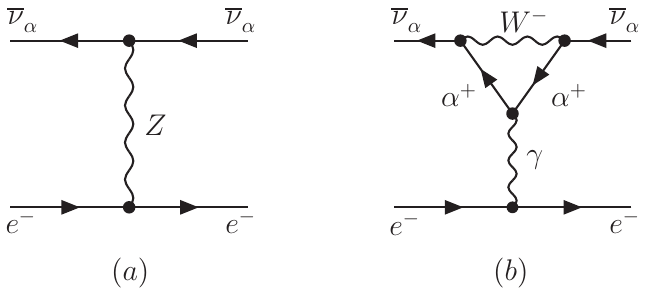}
\vspace{-0.2cm}
\caption{The Feynman diagrams of elastic $\overline{\nu}^{}_\alpha$-$e^-$ scattering 
(for $\alpha = \mu, \tau$) via weak neutral-current interactions: (a) the tree-level
contribution; (b) the dominant one-loop contribution, where only one typical diagram 
is shown for illustration. It is the difference between the charged-lepton masses
$m^{}_\mu$ and $m^{}_\tau$ that makes the $\overline{\nu}^{}_\mu$-$e^-$ and 
$\overline{\nu}^{}_\tau$-$e^-$ scattering cross sections distinguishable. Note that 
a full set of the one-loop Feynman diagrams will be taken into account in our 
analytical and numerical calculations of the $\overline{\nu}^{}_\alpha$-$e^-$ 
scattering processes.}
\label{fig:Feynman}
\end{figure}

\section{Elastic scattering}

The elastic scattering of $\overline{\nu}^{}_\mu$ or $\overline{\nu}^{}_\tau$
with an electron takes place via the standard weak neutral-current interactions 
as typically illustrated in Fig.~\ref{fig:Feynman},
whereas that of $\overline{\nu}^{}_e$ with an electron occurs through both
neutral- and charged-current interactions. After all the one-loop radiative 
corrections have been taken into account~\cite{Marciano:1980pb,Sarantakos:1982bp,
Huang:2024rfb}, the differential cross sections of elastic 
$\overline{\nu}^{}_e$-$e^-$, $\overline{\nu}^{}_\mu$-$e^-$ and 
$\overline{\nu}^{}_\tau$-$e^-$ scattering reactions are found to be
\begin{eqnarray}
\frac{{\rm d}\sigma^{}_e}{{\rm d}T^{}_e} & = &
\frac{2 G_\mu^2 \rho^2 m_e^{}}{\pi}\left[\kappa^2_e s^4_{\rm W}
+ \left(1-\frac{T^{}_e}{E}\right)^2 \left(\frac{\rho - 2}{2\rho}
- \kappa^{}_e s^2_{\rm W}\right)^2 + \frac{m_e^{} T^{}_e}{E^{2}}
\kappa^{}_e s^2_{\rm W} \left(\frac{\rho - 2}{2\rho}
- \kappa^{}_e s^2_{\rm W}\right)\right] \; ,
\nonumber \\
\frac{{\rm d}\sigma^{}_\mu}{{\rm d}T^{}_e} & = &
\frac{2 G_\mu^2 \rho^2 m_e^{}}{\pi}
\left[\kappa^2_\mu s^4_{\rm W} + \left(1 - \frac{T^{}_e}{E}\right)^2
\left(\frac{1}{2} - \kappa^{}_\mu s^2_{\rm W}\right)^2
+ \frac{m_e^{} T^{}_e}{E^{2}} \kappa^{}_\mu s^2_{\rm W}
\left(\frac{1}{2} - \kappa^{}_\mu s^2_{\rm W}\right)\right] \; ,
\nonumber \\
\frac{{\rm d}\sigma^{}_\tau}{{\rm d}T^{}_e} & = &
\frac{2 G_\mu^2 \rho^2 m_e^{}}{\pi}
\left[\kappa^2_\tau s^4_{\rm W} + \left(1 - \frac{T^{}_e}{E}\right)^2
\left(\frac{1}{2} - \kappa^{}_\tau s^2_{\rm W}\right)^2
+ \frac{m_e^{} T^{}_e}{E^{2}} \kappa^{}_\tau s^2_{\rm W}
\left(\frac{1}{2} - \kappa^{}_\tau s^2_{\rm W}\right)\right] \; ,
\label{6}
\end{eqnarray}
where $T^{}_e \equiv E^{}_e - m^{}_e$ stands for the recoil energy of the final-state
electron, $G^{}_\mu$ is the Fermi coupling constant determined from a precision
measurement of the muon lifetime, $m^{}_e$ denotes the electron mass, $E$ represents
the initial neutrino energy, and $s^{}_{\rm W} \equiv \sin\theta^{}_{\rm W}$ with
$\theta^{}_{\rm W}$ being the Weinberg angle. Moreover, the overall factor $\rho$ in
Eq.~(\ref{6}) is given by
\begin{eqnarray}
\rho & = & 1 + \frac{\alpha^{}_{\rm em}}{4\pi}
\left\{\frac{3}{4s^2_{\rm W}} \ln c^2_{\rm W} - \frac{7}{4s^2_{\rm W}}
+ \frac{2 c_Z^{}}{c^2_{\rm W} s^2_{\rm W}}
+ G \left(c^2_{\rm W} , \frac{m_h^2}{m_Z^2}\right) + \frac{1}{2 m_W^2
s^2_{\rm W}} \sum_f m_f^2 \ln\left(\frac{m_f^2}{m_W^2}\right) \right.
\nonumber \\
& & - \hspace{0.05cm} 2 \sum_{q, \hspace{0.05cm} q^\prime}
\left|V_{qq^\prime}^{}\right|^2 \int_{0}^{1} {\rm d} x
\Big[m_q^2 x + m_{q^\prime}^2 \left(1 - x\right)\Big] \ln\left[\frac{m_q^2 x
+ m_{q^\prime}^2 \left(1-x\right)}{m_W^2}\right]
\nonumber \\
& & \left. - \hspace{0.05cm} 2 \sum_{\alpha} \int_{0}^{1} {\rm d}x
\Big[m_{\alpha}^2 \left(1-x\right)\Big]
\ln\left[\frac{ m_{\alpha}^2 \left(1-x\right)}{m_W^2}\right] \right\} \; ,
\label{7}
\end{eqnarray}
where $\alpha^{}_{\rm em}$ denotes the electromagnetic fine-structure constant,
$m^{}_h$, $m^{}_W$ and $m^{}_Z$ stand respectively for the Higgs-, $W$- and $Z$-boson
masses, $m^{}_f$ represents the charged-fermion mass, $m^{}_\alpha$ refers to the
charged-lepton mass, $m^{}_q$ and $m^{}_{q^\prime}$
are the respective up- and down-type quark masses, $c^{}_{\rm W} \equiv
\cos\theta^{}_{\rm W}$, $V^{}_{q q^\prime}$ denote the elements of the
Cabibbo-Kobayashi-Maskawa (CKM) quark flavor mixing matrix (for $q = u, c, t$ and
$q^\prime = d, s, b$), the coefficient $c^{}_Z$ and the function $G(c^2_{\rm W} , x)$
are defined as
\begin{eqnarray}
c_Z^{} \equiv \frac{19}{8} - \frac{7}{2} s^2_{\rm W} + 3 s^4_{\rm W} \; , \quad
G\left(c^2_{\rm W} , x\right) \equiv \frac{3}{4} \cdot \frac{x}{s^2_{\rm W}}
\left[\frac{1}{c^2_{\rm W} - x} \ln\left(\frac{c^2_{\rm W}}{x}\right)
+ \frac{1}{c^2_{\rm W}\left(1 - x\right)} \ln x \right] \; ,
\label{8}
\end{eqnarray}
and the summation over $q$ and $q^\prime$ quarks in Eq.~(\ref{7}) should include their
color factor ``3". On the other hand, the flavor-dependent functions $\kappa^{}_\alpha$
(for $\alpha = e, \mu, \tau$) are expressed as
\begin{eqnarray}
\kappa_\alpha\left(q^2\right) & = & 1 - \left\{\frac{c^{}_{\rm W}}{s^{}_{\rm W}}
\cdot \frac{\Sigma_{AZ}^{} \left(q^2\right)}{q^2} + \frac{c^2_{\rm W}}{s^2_{\rm W}}
{\rm Re}\left[\frac{\Sigma_Z^{}\left(m_Z^2\right)}{m_Z^2}
- \frac{\Sigma_W^{}\left(m_W^2\right)}{m_W^2}\right]
- \frac{\alpha^{}_{\rm em}}{2\pi} \cdot \frac{c^2_{\rm W}}{s^2_{\rm W}}
\left[\left(\frac{m_Z^2}{q^2} - 2\right) \right.\right.
\nonumber \\
& & \times \left.\left.
\left[ \frac{1}{\epsilon} - \gamma^{}_{\rm E}
+ \ln 4\pi + \ln\left(\frac{\mu^2}{m_W^2}\right)\right] \right]
- \frac{1}{c^2_{\rm W}} \left[\frac{c_A^{}}{c^2_{\rm W}}
- R^{}_\alpha\left(q^2\right)\right]\right\} \; ,
\label{9}
\end{eqnarray}
where $q^2 = -2m^{}_e T^{}_e$ is the square of momentum transfer, and
\begin{eqnarray}
c_A^{} \equiv \frac{19}{8} - \frac{17}{4} s^2_{\rm W} + 3s^4_{\rm W} \; ,
\quad
R^{}_\alpha\left(q^2\right) \equiv \frac{1}{3} + 2 \int_{0}^{1} {\rm d}x
\Big[ x\left(1-x\right)\Big] \ln\left[\frac{m_W^2}{m_\alpha^2
- q^2 x \left(1-x\right)}\right] \; .
\label{10}
\end{eqnarray}
In Eq.~(\ref{9}) the ultra-violet divergence term and the $\mu$-dependence
term stemming from dimensional regularization (with $\mu$ being the 't Hooft
mass scale, $d \equiv 4 - 2\epsilon$ being the spacetime dimension and
$\gamma^{}_{\rm E} \simeq 0.577$ being the Euler-Mascheroni constant) will
cancel the divergence terms associated with the one-loop self-energy functions
of the gauge bosons $\Sigma_{AZ}^{} (q^2)$, $\Sigma_{W}^{} (m_W^2)$ and
$\Sigma_{Z}^{} (m_Z^2)$~\cite{Marciano:1980pb}.

It is obvious that switching off the one-loop radiative corrections is
equivalent to taking $\rho = 1$ and
$\kappa^{}_e = \kappa^{}_\mu = \kappa^{}_\tau = 1$, from which the
tree-level equality $\sigma^{}_\mu = \sigma^{}_\tau$ is simply obtained.
Note that $R^{}_\mu(q^2)$ and $R^{}_\tau(q^2)$ arise respectively from the
one-loop $\overline{\nu}^{}_\mu$-$\overline{\nu}^{}_\mu$-$\gamma$ and
$\overline{\nu}^{}_\tau$-$\overline{\nu}^{}_\tau$-$\gamma$ vertices as 
typically illustrated by Fig.~\ref{fig:Feynman} ($b$), and their 
discrepancy is attributed to the difference between $m^2_\mu$ and
$m^2_\tau$ as can be easily seen from Eq.~(\ref{10}). If the resulting
discrepancy between elastic $\overline{\nu}^{}_\mu$-$e^-$ and
$\overline{\nu}^{}_\tau$-$e^-$ scattering cross sections is observed from
a reactor antineutrino oscillation experiment, it will allows us to probe or
constrain the fine effects of leptonic CP violation and $\mu$-$\tau$ interchange
symmetry breaking as described by $P^{}_-$ in Eq.~(\ref{4}).

Let us proceed to calculate the rates of $\overline{\nu}^{}_e$,
$\overline{\nu}^{}_\mu$ and $\overline{\nu}^{}_\tau$ events at the far
detector. The initial reactor electron antineutrinos are produced from
the nuclear beta decays, and the fluxes of $\overline{\nu}^{}_\alpha$ events
originating from $\overline{\nu}^{}_e \to \overline{\nu}^{}_\alpha$ oscillations
(for $\alpha = e, \mu, \tau$) with a baseline length $L$ are given as
\begin{eqnarray}
\phi^{}_\alpha\left(E\right) = \frac{1}{4\pi L^2} \cdot P(\overline{\nu}^{}_e
\to \overline{\nu}^{}_\alpha) \cdot \frac{{\rm d}N^{\prime}_e}{{\rm d}E} \; ,
\label{11}
\end{eqnarray}
where ${\rm d}N^\prime_e/{\rm d}E$ is the rate for the initial
$\overline{\nu}^{}_e$ events. In the far detector, the event rates of elastic
$\overline{\nu}^{}_\alpha + e^- \to \overline{\nu}^{}_\alpha + e^-$ scattering
turn out to be
\begin{eqnarray}
\frac{{\rm d}N^{}_{\alpha}}{{\rm d}T^{\prime}_e} = N^{}_{\rm T} T
\int^{\infty}_{0} {\rm d}T^{}_e \hspace{0.05cm}
\frac{1}{\sqrt{2\pi} \hspace{0.05cm} \delta T^{}_e}
\exp\left[-\frac{\left(T^{}_e - T^{\prime}_e\right)^2}{2\left(\delta T^{}_e\right)^2}
\right] \int^\infty_{E^{}_{\rm min}} {\rm d}E \hspace{0.05cm}
\frac{{\rm d}\sigma^{}_\alpha}{{\rm d}T^{}_e} \hspace{0.05cm} \phi^{}_\alpha \; ,
\label{12}
\end{eqnarray}
where $N^{}_{\rm T}$ is the total number of the target electrons in the far detector,
$T$ denotes the operation duration, $T^\prime_e$ represents the observed energy in
the detector corresponding to the electron recoil energy $T^{}_e$ and the energy
resolution $\delta T^{}_e$ with a Gaussian distribution, the differential cross
sections ${\rm d}\sigma^{}_\alpha/{\rm d}T^{}_e$ and the fluxes $\phi^{}_\alpha$
have been given in Eqs.~(\ref{6}) and (\ref{11}) respectively.

To see the flavor effects in a clear way, we rewrite the integrand of the second
integral in Eq.~(\ref{12}) and sum over the flavor index $\alpha$. Then we are left
with
\begin{eqnarray}
\sum_\alpha \frac{{\rm d}\sigma^{}_\alpha}{{\rm d}T^{}_e} \hspace{0.05cm} \phi^{}_\alpha
= \frac{1}{4\pi L^2} \cdot \frac{{\rm d}N^\prime_e}{{\rm d}E}
\left[P(\overline{\nu}^{}_e \to \overline{\nu}^{}_e) \hspace{0.05cm}
\frac{{\rm d}\sigma^{}_e}{{\rm d}T^{}_e} +
\frac{P^{}_+}{2} \left(\frac{{\rm d}\sigma^{}_\mu}{{\rm d}T^{}_e} +
\frac{{\rm d}\sigma^{}_\tau}{{\rm d}T^{}_e}\right) +
\frac{P^{}_-}{2} \left(\frac{{\rm d}\sigma^{}_\mu}{{\rm d}T^{}_e} -
\frac{{\rm d}\sigma^{}_\tau}{{\rm d}T^{}_e}\right)\right] \; ,
\label{13}
\end{eqnarray}
where the expressions of $P^{}_\pm$ have been given in Eq.~(\ref{4}). Since the
$\overline{\nu}^{}_e$ contribution can be subtracted from the total elastic
scattering events by accurately measuring the $\overline{\nu}^{}_e$ flux via the
inverse beta decay $\overline{\nu}^{}_e + p \to e^+ + n$, we are therefore able
to observe the appearance of $\overline{\nu}^{}_\mu$ and $\overline{\nu}^{}_\tau$
events as a whole. In other words, it is relatively easy to extract $P^{}_+$ from
such a precision measurement, and extracting the information on $P^{}_-$ will be
very challenging. But the latter is highly nontrivial and deserves to be
carefully studied, simply because it may open a new window to probe or constrain
leptonic CP violation in the reactor-based experiments.

\begin{figure}[t!]
\centering
\includegraphics[width=\textwidth]{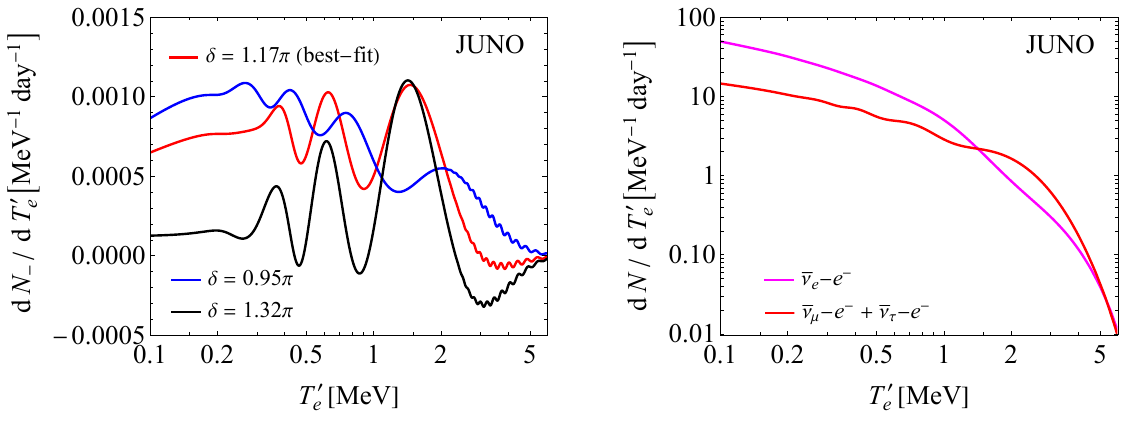}
\vspace{-0.8cm}
\caption{In the left panel, the $\delta$-dependent contribution 
$P^{}_- \left( {\rm d}\sigma^{}_\mu/{\rm d}T^{}_e - 
{\rm d}\sigma^{}_\tau/{\rm d}T^{}_e\right)/2$ to the event rate is denoted as 
${\rm d}N^{}_-/{\rm d}T^\prime_e$, and its numerical
result is illustrated by inputting the best-fit value $\delta = 1.17\pi$ and 
its $1\sigma$ lower (upper) bound $\delta = 0.95\pi$ ($1.32\pi$).
In the right panel, the event rate ${\rm d}N^{}_e/{\rm d}T^\prime_e$ for $\overline{\nu}^{}_e$-$e^-$ scattering and the sum 
${\rm d}N^{}_\mu/{\rm d}T^\prime_e + {\rm d}N^{}_\tau/{\rm d}T^\prime_e$ of 
the event rates for $\overline{\nu}^{}_\mu$-$e^-$ and 
$\overline{\nu}^{}_\tau$-$e^-$ scattering are presented. Here the JUNO detector 
with 20 kiloton liquid scintillator and an
effective energy resolution $\delta T^{}_e/T^{}_e = 3\%/\sqrt{T^{}_e/{\rm MeV}}$
have been taken into consideration.}
\label{fig:events}
\end{figure}

To illustrate, let us specify the nominal setup for a reactor antineutrino experiment
by taking the JUNO detector for example, and calculate the elastic
$\overline{\nu}^{}_\alpha$-$e^-$ scattering event rates (for $\alpha = e, \mu, \tau$).
Given the total thermal power $P^{}_{\rm th}$ of the chosen nuclear reactors,
the production rate of the reactor electron antineutrinos can be calculated by use of
\begin{eqnarray}
\frac{{\rm d}N^\prime_e}{{\rm d}E} = \frac{P^{}_{\rm th}}{\displaystyle
\sum_i f^{}_i \epsilon^{}_i} \sum_i f^{}_i S^{}_{i}(E) \; ,
\label{14}
\end{eqnarray}
where the subscript ``$i$" refers to the radioactive isotopes ${^{235}{\rm U}}$,
${^{238}{\rm U}}$, ${^{239}{\rm Pu}}$ and ${^{241}{\rm Pu}}$, $f^{}_i$ stands for
the average fraction of each isotope, and $\epsilon^{}_i$ denotes the thermal energy
released per fission. The energy spectrum $S^{}_i(E)$ of $\overline{\nu}^{}_e$ from
the beta decays of fission products is obtained with the help of a fifth-power
fitting formula~\cite{Mueller:2011nm}. We assume
$P^{}_{\rm th} = 26.6~{\rm GW}$, $f^{}_i = \{0.561, 0.076, 0.307, 0.056\}$ and
$\epsilon^{}_i = \{202.36, 205.99, 211.12, 214.26\}~{\rm MeV}$ as the typical
inputs in our numerical calculation
\footnote{Note that
$1~{\rm GW} = 6.2415 \times 10^{21}~{\rm MeV}\cdot {\rm s}^{-1}$ holds.},
and use the values of $\alpha^{}_{ij}$ from Ref.~\cite{Mueller:2011nm}. Then
the elastic scattering event rates at the far detector can be calculated by
convolving the antineutrino fluxes and the differential cross sections,
given the total number of target electrons
$N^{}_{\rm T} \simeq 6.72\times 10^{33}$ corresponding to the $12\%$ mass fraction
of hydrogen and $88\%$ of carbon in the 20 kiloton liquid scintillator. We show the 
final numerical results in Fig.~\ref{fig:events}. Some comments are in order.
\begin{itemize}
  \item In the left panel, the contribution to the event rate in Eq.~(\ref{12}) 
  from the last term on the right-hand side of Eq.~(\ref{13}) has been illustrated, 
  where the dependence on the CP-violating phase $\delta$ is evident. For simplicity, 
  we take the best-fit value $\delta = 1.17\pi$ and its $1\sigma$ lower (upper) 
  bound $\delta = 0.95\pi$ ($1.32\pi$) from Ref.~\cite{Esteban:2024eli}. Note that 
  this best-fit result agrees perfectly with $\delta = 1.20\pi$ obtained from Ref.~\cite{Capozzi:2025wyn} in the case of normal neutrino mass ordering. Although 
  ${\rm d}N^{}_-/{\rm d}T^\prime_e$ is very sensitive to the values of $\delta$, 
  the variation of the event rates at the JUNO turns out to be $0.001~{\rm MeV}^{-1} 
  {\rm day}^{-1}$ at most. As indicated in Eq.~(\ref{13}), the suppression arises 
  mainly from the difference between the one-loop cross sections for $\overline{\nu}^{}_\mu$-$e^-$ and $\overline{\nu}^{}_\tau$-$e^-$ scattering. 
  The ratio of such a difference to the tree-level cross section is about $1\%$, 
  while the oscillation probability difference $P^{}_-$ contributes some extra 
  suppression. An experimental determination of the event rate at the level of 
  $0.001~{\rm MeV}^{-1} {\rm day}^{-1}$ is no doubt a big challenge.

  \item In the right panel, the sum ${\rm d}N^{}_\mu/{\rm d}T^\prime_e + 
  {\rm d}N^{}_\tau/{\rm d}T^\prime_e$ over the event rates for 
  $\overline{\nu}^{}_\mu$-$e^-$ and $\overline{\nu}^{}_\tau$-$e^-$ scattering is 
  presented together with ${\rm d}N^{}_e/{\rm d}T^\prime_e$ for 
  $\overline{\nu}^{}_e$-$e^-$ scattering. Below $T^\prime_e \approx 1.5~{\rm MeV}$, 
  the $\overline{\nu}^{}_e$ event rate is always dominant. However, for 
  $T^\prime_e > 1.5~{\rm MeV}$, we have more $\overline{\nu}^{}_\mu$ and $\overline{\nu}^{}_\tau$ events. This observation can be easily understood as 
  follows. The oscillation probability $P^{}_+$ is larger than 
  $P(\overline{\nu}^{}_e \to \overline{\nu}^{}_e)$ for the antineutrino beam energy 
  $E \in [2, 6]~{\rm MeV}$, indicating that more $\overline{\nu}^{}_\mu$ and $\overline{\nu}^{}_\tau$ events than $\overline{\nu}^{}_e$ arrive in the far 
  detector. As mentioned before, the $\overline{\nu}^{}_e$ flux will be precisely 
  measured through the inverse beta decay such that its contribution to the elastic 
  scattering events can be reliably subtracted. For the visible energy 
  $T^\prime_e \in [0.1, 4]~{\rm MeV}$, the appearance of $\overline{\nu}^{}_\mu$ 
  and $\overline{\nu}^{}_\tau$ at the JUNO could be optimistically demonstrated by 
  observing the elastic scattering events at the rate of 
  $(1\cdots 10)~{\rm MeV}^{-1}~{\rm day}^{-1}$.
\end{itemize}
Within the energy region $T^\prime_e \in [0.1, 4]~{\rm MeV}$, a number of 
backgrounds for observing the recoiled electrons should be taken into account, 
including the internal background due to the radioactivity of contaminants in the 
scintillator, the external backgrounds from the $\gamma$ radioactivity of 
surrounding materials and beta decays of cosmogenic isotopes from the spallation 
of cosmic muons. According to the background analysis for the detection of solar 
neutrinos at the  JUNO~\cite{JUNO:2023zty}, in an ideal radiopurity scenario, the 
largest internal background comes from the $^{210}{\rm Pb}$ chain and the 
$^{238}{\rm U}$ chain with a total rate about $419$ counts per kiloton per day in 
the range $T^\prime_e \in [0.45, 1.6]~{\rm MeV}$. The external $\gamma$ background 
can be eliminated by only retaining the events within the spherical fiducial volume 
of a radius about $15$ meters. The cosmogenic background is dominated by the beta 
decays of $^{11}{\rm C}$ at a rate of about $1761$ counts per kiloton per day. In 
addition, the solar neutrinos themselves serve as an irreducible background, where 
the dominant contribution of $500$ counts per kiloton per day is made by 
$^{7}{\rm Be}$ neutrinos. Obviously, the background rate is much larger than the 
signal rate associated with the flavor oscillations of reactor antineutrinos. One 
possible way out is to implement the directional information on the recoiled 
electrons such that the antineutrinos coming from the direction opposite to the 
nuclear reactors can be distinguished from the others.

Furthermore, the dependence on the leptonic CP-violating phase $\delta$ at the 
per-mille level will be relevant if the statistical uncertainty in the total number 
of $\overline{\nu}^{}_\mu$-$e^-$ and $\overline{\nu}^{}_\tau$-$e^-$ scattering 
events reaches the same level, i.e., with $10^6$ events. This can be realized by 
increasing the thermal power of reactors, optimizing the baseline in order to 
enhance the CP-violating effect and the antineutrino fluxes, and even constructing 
a much larger far detector. 

\section{Summary}

We have carried out a {\it proof-of-concept} study of the possibility to observe 
the appearance of $\overline{\nu}^{}_\mu$ and $\overline{\nu}^{}_\tau$ events in 
the far detector of a reactor antineutrino oscillation experiment. Taking the JUNO 
experiment for example, we have calculated the differential event rate for $\overline{\nu}^{}_\mu$-$e^-$ and $\overline{\nu}^{}_\tau$-$e^-$ scattering and 
find that the total rate is about $9$ counts per day for a 20 kiloton 
liquid-scintillator detector, while that for $\overline{\nu}^{}_e$-$e^-$ scattering 
is about $17$ counts per day. However, the $\overline{\nu}^{}_e$ events can be 
eliminated by precisely measuring the $\overline{\nu}^{}_e$ flux through the inverse 
beta decay $\overline{\nu}^{}_e + p \to e^+ + n$. We stress that it is important to experimentally verify the appearance of $\overline{\nu}^{}_\mu$ and 
$\overline{\nu}^{}_\tau$ events, so as to test the conservation of probability
for reactor antineutrino oscillations.

When the one-loop correction to the cross sections for antineutrino-electron 
scattering is taken into account, the event rate becomes dependent on the leptonic 
CP-violating phase $\delta$ through the oscillation probability difference 
$P^{}_- \equiv P(\overline{\nu}^{}_e \to \overline{\nu}^{}_\mu) - 
P(\overline{\nu}^{}_e \to \overline{\nu}^{}_\tau)$. Such dependence relies on both 
a nonzero $P^{}_-$ and the difference in the cross sections of elastic $\overline{\nu}^{}_\mu$-$e^-$ and $\overline{\nu}^{}_\tau$-$e^-$ scattering, 
and hence it is highly suppressed. That is why we expect that an unprecedented 
precision measurement of elastic antineutrino-electron scattering events at the 
per-mille level is required to probe or constrain leptonic CP violation in the 
foreseeable future.

\section*{Acknowledgments}

The authors are indebted to Jihong Huang for his kind technical helps and valuable 
discussions. This work was supported in part by the National Natural Science 
Foundation of China under grant No. 12475113, the CAS Project for Young Scientists 
in Basic Research (YSBR-099), and the Scientific and Technological Innovation 
Program of IHEP under grant No. E55457U2.

\end{document}